\documentclass[3p,sort&compress,longtitle]{elsarticle}
\usepackage[utf8]{inputenc}
\usepackage{xcolor}
\usepackage{lmodern}
\usepackage[english]{babel}
\usepackage{siunitx} 
\usepackage{booktabs}
\usepackage{comment}
\usepackage[pdftex]{hyperref}
\hypersetup{%
   pdfauthor=Simone Valdré,%
   pdftitle=The FAZIA setup: a review on the electronics and the mechanical mounting%
}

\journal{Nuclear Instruments and Methods in Physics Research A}

\begin{document}
	\begin{frontmatter}
		
		\title{The FAZIA setup: a review on the electronics\\and the mechanical mounting\tnoteref{cr}}
		
		\author[infn-fi]{S.~Valdré\corref{valdre}}
		\cortext[valdre]{Corresponding author}
		\ead{valdre@fi.infn.it}
		\author[infn-fi]{G.~Casini}
		\author[lpc]{N.~Le Neindre}
		\author[unifi,infn-fi]{M.~Bini\fnref{retired}}
		\author[infn-na]{A.~Boiano}
		\author[ipno]{B.~Borderie}
		\author[ipno]{P.~Edelbruck\fnref{retired}}
		\author[unifi,infn-fi]{G.~Poggi}
		\author[ipno]{F.~Salomon}
		\author[infn-na]{G.~Tortone}
		
		\author[infn-lns]{R.~Alba}
		\author[unifi,infn-fi]{S.~Barlini}
		\author[subatech]{E.~Bonnet}
		\author[lpc]{R.~Bougault}
		\author[lpc]{B.~Bougard}
		\author[ipno]{G.~Brulin}
		\author[unibo,infn-bo]{M.~Bruno\fnref{retired}}
		\author[unifi,infn-fi]{A.~Buccola}
		\author[unifi,infn-fi]{A.~Camaiani}
		\author[ganil]{A.~Chbihi}
		\author[unifi,infn-fi]{C.~Ciampi}
		\author[unipd,infn-lnl]{M.~Cicerchia}
		\author[infn-lnl]{M.~Cinausero}
		\author[msu]{D.~Dell'Aquila}
		\author[lpc]{P.~Desrues}
		\author[huelva]{J.~A.~Due\~{n}as}
		\author[infn-pd]{D.~Fabris}
		\author[unifi,infn-fi]{M.~Falorsi\fnref{retired}}
		\author[ganil]{J.~D.~Frankland}
		\author[unifi,infn-fi]{C.~Frosin}
		\author[ipno,arts]{E.~Galichet}
		\author[unina]{R.~Giordano}
		\author[infn-lnl]{F.~Gramegna}
		\author[ipno]{L.~Grassi}
		\author[lpc]{D.~Gruyer}
		\author[infn-bo]{M.~Guerzoni}
		\author[lpc]{M.~Henri}
		\author[nowoc]{M.~Kajetanowicz}
		\author[ifj]{K.~Korcyl}
		\author[warsaw]{A.~Kordyasz}
		\author[krakow]{T.~Kozik}
		\author[ganil]{P.~Lecomte}
		\author[infn-ct]{I.~Lombardo}
		\author[lpc]{O.~Lopez}
		\author[infn-lns]{C.~Maiolino}
		\author[unipd,infn-lnl]{G.~Mantovani}
		\author[infn-lnl]{T.~Marchi}
		\author[infn-bo]{A.~Margotti}
		\author[lpc]{Y.~Merrer}
		\author[ganil]{L.~Morelli}
		\author[infn-fi]{A.~Olmi\fnref{retired}}
		\author[infn-na]{A.~Ordine}
		\author[unifi,infn-fi]{P.~Ottanelli}
		\author[lpc]{C.~Pain}
		\author[krakow]{M.~Pałka}
		\author[lpc,hulubei]{M.~P\^{a}rlog}
		\author[unifi,infn-fi]{G.~Pasquali}
		\author[unifi,infn-fi]{G.~Pastore}
		\author[infn-fi]{S.~Piantelli}
		\author[lpc]{H.~de Préaumont}
		\author[ganil]{R.~Revenko}
		\author[ipno]{A.~Richard\fnref{retired}}
		\author[ipno]{M.F.~Rivet\fnref{deceased}}
		\author[ganil]{J.~Ropert}
		\author[unina,infn-na]{E.~Rosato\fnref{deceased}}
		\author[ganil]{F.~Saillant}
		\author[infn-lns]{D.~Santonocito}
		\author[unifi,infn-fi]{E.~Scarlini}
		\author[infn-bo]{S.~Serra}
		\author[ipno]{C.~Soulet}
		\author[unina,infn-na]{G.~Spadaccini\fnref{retired}}
		\author[unifi,infn-fi]{A.~A.~Stefanini}
		\author[infn-fi]{G.~Tobia}
		\author[krakow]{S.~Upadhyaya}
		\author[infn-na]{A.~Vanzanella}
		\author[infn-ct]{G.~Verde}
		\author[lpc]{E.~Vient}
		\author[unina,infn-na]{M.~Vigilante}
		\author[ipno]{E.~Wanlin}
		\author[ganil]{G.~Wittwer}
		\author[infn-bo]{A.~Zucchini}
		
		\address[infn-fi]{INFN --- Sezione di Firenze, 50019 Sesto Fiorentino, Italy}
		\address[lpc]{LPC Caen, Normandie Univ, ENSICAEN, UNICAEN, CNRS/IN2P3, LPC Caen, 14000 Caen, France}
		\address[unifi]{Dipartimento di Fisica, Università di Firenze, 50019 Sesto Fiorentino, Italy}
		\address[infn-na]{INFN --- Sezione di Napoli, 80126 Napoli, Italy}
		\address[ipno]{Institut de Physique Nucléaire, CNRS-IN2P3, Univ. Paris-Sud, Université Paris-Saclay, 91406 Orsay, France}
		\address[infn-lns]{INFN --- Laboratori Nazionali del Sud, 95123 Catania, Italy}
		\address[subatech]{SUBATECH, EMN-IN2P3/CNRS-Université de Nantes, Nantes, France}
		\address[unibo]{Dipartimento di Fisica, Università di Bologna, 40127 Bologna, Italy}
		\address[infn-bo]{INFN --- Sezione di Bologna, 40127 Bologna, Italy}
		\address[ganil]{GANIL, CEA/DRF-CNRS/IN2P3, 14076 Caen, France}
		\address[unipd]{Dipartimento di Fisica, Università di Padova, 35131 Padova, Italy}
		\address[infn-lnl]{INFN --- Laboratori Nazionali di Legnaro, 35020 Legnaro, Italy}
		\address[msu]{National Superconducting Cyclotron Laboratory, Michigan State University, East Lansing, Michigan 48824, USA}
		\address[huelva]{Depto. de Ingenier\'ia El\'ectrica y Centro de Estudios Avanzados en F\'isica, Matem\'aticas y Computaci\'on. Universidad de Huelva, 21071 Huelva, Spain}
		\address[infn-pd]{INFN --- Sezione di Padova, 35131 Padova, Italy}
		\address[arts]{Conservatoire National des Arts et Métiers, 75141 Paris cedex 03, France}
		\address[unina]{Dipartimento di Fisica, Università di Napoli, 80126 Napoli, Italy}
		\address[nowoc]{Nowoczesna Elektronika, 30-109 Cracow, Poland}
		\address[ifj]{Institute of Nuclear Physics, Polish Academy of Sciences, 31-342 Cracow, Poland}
		\address[warsaw]{Heavy Ion Laboratory, University of Warsaw, 02-093 Warszawa, Poland}
		\address[krakow]{Faculty of Physics, Astronomy and Applied Computer Science, Jagiellonian University, 30-348 Cracow, Poland}
		\address[infn-ct]{INFN --- Sezione di Catania, 95123 Catania, Italy}
		\address[hulubei]{``Horia Hulubei'' National Institute for R\&D in Physics and Nuclear Engineering (IFIN-HH), P.O.BOX MG-6, Bucharest Magurele, Romania}
		
		\tnotetext[cr]{\textcopyright\ 2018. This manuscript version is made available under the CC-BY-NC-ND 4.0 license.\\\url{http://creativecommons.org/licenses/by-nc-nd/4.0/}}
		\fntext[retired]{Retired}
		\fntext[deceased]{Deceased}
		
		\begin{abstract}
			In this paper the technological aspects of the FAZIA array will be explored. After a productive commissioning phase,
			FAZIA blocks started to measure and give very useful data to explore the physics of Fermi energy heavy-ion reactions.
			This was possible thanks to many technical measures and innovations developed in the commissioning phase
			and tuned during the first experimental campaigns. This paper gives a detailed description of the present status of the FAZIA setup
			from the electronic and mechanical point of view, trying also to trace a path for new improvements and refinements of the apparatus.
		\end{abstract}
		
		
	\end{frontmatter}
	
	\section{Introduction and specific requirements}
	FAZIA (\textit{Forward-angle A and Z Identification Array}) is a modern and innovative three-layer telescope [Si + Si + CsI(Tl)] array.
	This review paper follows the work of R. Bougault \textit{et al.} \cite{Bougault14} where the R\&D phase
	of the FAZIA project was carefully described.
	In this paper we will give a detailed overview on the technological aspects of the final detector array;
	in particular, we will focus on the electronics, the data flow and the mechanical aspects of the apparatus.
	
	The multi-detector FAZIA aims at detecting and identifying particles and fragments produced
	in heavy-ion reactions around Fermi energy. The main requirement of FAZIA is the modularity
	and portability: in fact, FAZIA was designed to measure in various laboratories,
	in different setups and coupled to several detectors. Another important objective is to
	maximise unit identification for charges and masses of detected nuclei.
	In the present situation, we clearly discriminate charges up to $Z\sim 55$ and masses up to $Z\sim 25$. This goal
	was achieved using custom detectors produced following a well-studied recipe \cite{VonAmmon92,Bardelli09}
	and using original electronics with novel \textit{pulse-shape discrimination} (PSD) techniques \cite{Ammerlaan63,Barlini09,Bardelli11} based
	on high speed analog-to-digital converters with rates up to \SI{250}{MS/s}
	and 14-bit resolution. The whole electronics is embedded in the proximity of the telescopes
	inside the vacuum chamber.
	
	The commissioning runs of FAZIA proved the capability to integrate inside the scattering
	chamber all the electronics required for silicon detectors and scintillators read-out by photodiodes.
	This very innovative electronics includes pre-amplifiers, analogue chains, high speed
	converters, read-out logic, high voltage devices and pulse generator for
	analogue chains. Indeed, the first experimental campaigns proved that it is possible to integrate, on the same
	multi-layer card, some potentially noisy power-supplies (such as switching regulators) with sensitive low-noise
	pre-amplifiers whose power-supply rejection ratio is not high. This integration has been possible by
	applying strict electromagnetic compatibility design guidelines.
	
	The great complexity and the fine granularity of the apparatus implies a difficult scalability and thus a relatively poor angular coverage:
	in fact a 16-telescope block, which is the smallest independent FAZIA unit, covers only around \SI{0.05}{\%} of the full solid angle at \SI{1}{m} distance from the target.
	Indeed, the next FAZIA schedule foresees the mounting of a 12-block array in 2019, coupled with INDRA detector array \cite{Pouthas95} at GANIL to increase the angular coverage.
	
	The electronic cards are described in Sec.~\ref{sec:electronics}.
	Both the ``block'' electronics, that is mounted directly next to detectors
	inside the scattering chamber (in vacuum) and the devices outside the chamber are detailed.
	Afterwards, a functional description of the apparatus follows (Sec.~\ref{sec:functional}): in particular
	the clock distribution (Sec.~\ref{ssc:clock}), the data packet structure (Sec.~\ref{ssc:packet}), the trigger logic (Sec.~\ref{ssc:trigger}), the data flow (Sec.~\ref{ssc:data}),
	the possibility of coupling (Sec.~\ref{ssc:coupling}), the acquisition (Sec.~\ref{ssc:acq}), and the slow control (Sec.~\ref{ssc:slow})
	are presented. In Sec.~\ref{sec:mechanics} the adopted mechanical solutions necessary
	to hold and to cool the block electronics and to align the detectors are described. Finally, in Sec.~\ref{sec:conclusions},
	we give an overview on all the innovative aspects of the FAZIA modular array and the next improvements we are going to realise.
	
	\section{Description of the electronic boards}\label{sec:electronics}
\begin{figure}[htbp]
		\centering
		\includegraphics[width=0.8\textwidth]{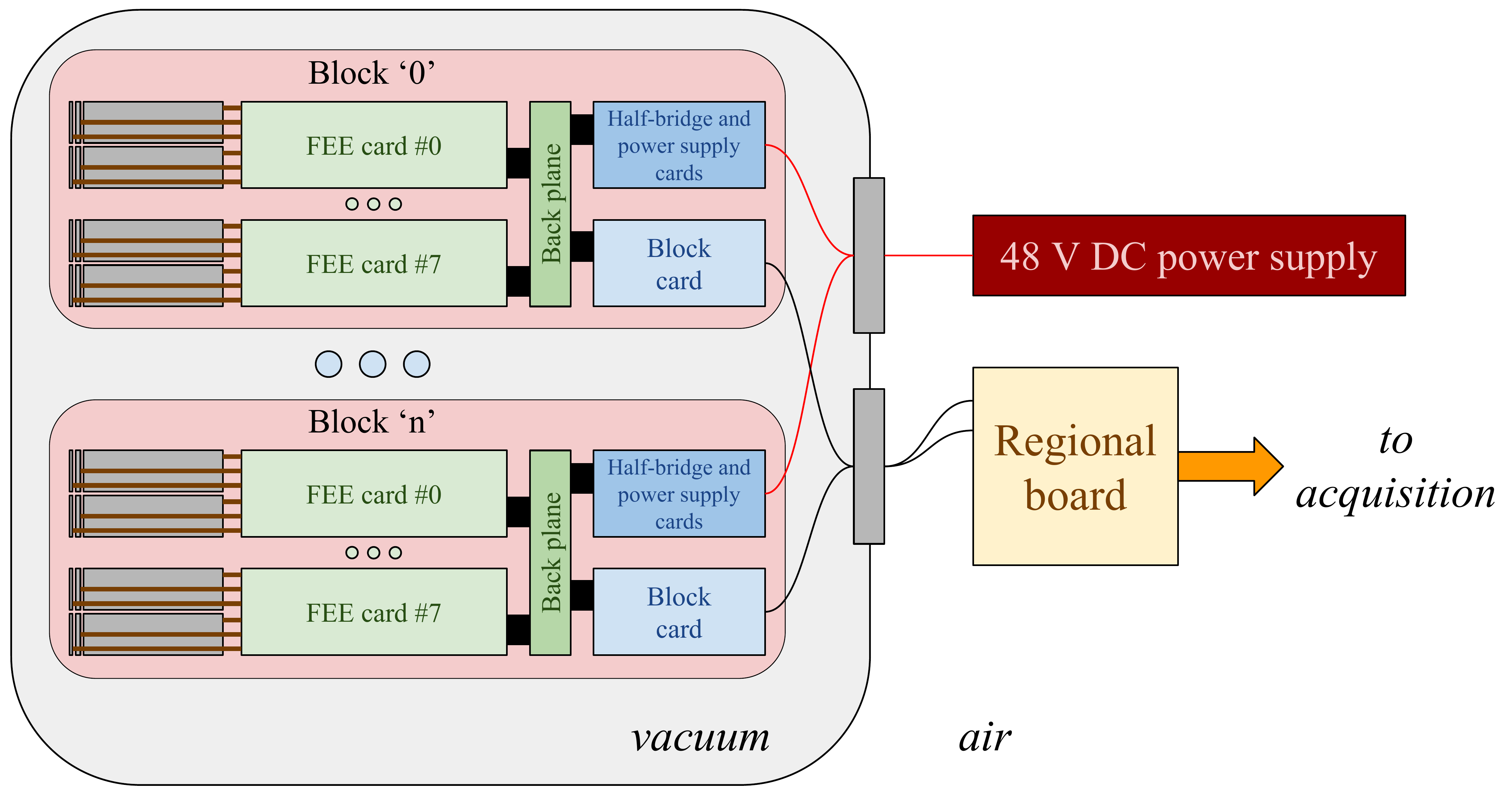}
		\caption{Schematic representation of the FAZIA electronic cards inside and outside the scattering chamber. For a detailed description see the text.}
		\label{fig:schema}
	\end{figure}

	The first feature which could be noticed when looking at the FAZIA apparatus
	is the absolute scarcity of electronic racks outside the scattering chamber (Fig.~\ref{fig:schema}).
	In fact, the so-called regional board (RB) is the only electronic card placed outside (see Sec.~\ref{ssc:rb})
	and it performs the functions of a standard ``event building'' card.
	Only two connections per block are necessary: a \SI{48}{V} (\SI{6}{A}) power supply line and a \SI{3}{Gbit/s} full-duplex
	optical link used to transfer data, to synchronise the clocks, to send triggers, and to manage any block parameter via slow control.
	
	Inside the vacuum chamber, the basic element of the FAZIA array is the block, which consists of 16 detector telescopes.
	The telescopes are connected to 8 front-end electronic (FEE) cards (see Sec.~\ref{ssc:fee}),
	which feature, among other components, charge sensitive pre-amplifiers, ADCs, Si bias voltage regulators, and FPGAs for data handling.
	Up to two telescopes can be connected to each front-end card.
	All the FEE cards are connected to a ``Block Card'' (see Sec.~\ref{ssc:bc}) via a common backplane, which hosts
	also two power supply boards (see Sec.~\ref{ssc:ps}). The Block Card (BC) is mainly devoted to handle I/O
	operations and to merge data coming from the FEE cards. Power Supply (PS) and Half Bridge (HB) cards produce and monitor the voltages
	needed to the other boards on the block.
	The FAZIA blocks communicate with the event building electronics via the Block Card
	through the \SI{3}{Gbit/s} optical link.
	A metallic housing covers all the block; there is also a metal screen around the BC, PS and HB cards to avoid that electromagnetic emissions
	from DC/DC converters reach the FEE cards. All the electronic cards of the block are supported by (and firmly screwed to) a copper plate in which water can flow for cooling.
	The FEE cards and the electromagnetic shield structure have been designed at IPN Orsay; BC, PS and HB boards have been developed at INFN -- Naples;
	the copper plates are built at INFN -- Bologna and Firenze; the RB has been designed at Jagiellonian University in Krak\'{o}w (Poland) in collaboration with INFN -- Naples.
	
	\subsection{Power supply}\label{ssc:ps}
	The DC levels needed by the block electronics (BC and FEEs) are produced by two cards that
	are both connected to the backplane: Half Bridge and Power Supply.
	As previously mentioned, the FAZIA block needs only an external \SI{48}{V} DC line to work. This voltage
	is provided to the Half Bridge card, which performs a high power conversion to \SI{22}{V} DC (\SI{14}{A} max) and \SI{5.5}{V} DC (\SI{70}{A} max)
	by means of two commercial switching converters (CUI VHB350 series). These devices have an insulation,
	with low parasitic capacity, between primary and secondary stage to avoid ground loops.
	To reduce the electromagnetic interferences produced by the converters, a particular two stage filter was
	added at the \SI{48}{V} input line: the first stage is a BALUN (\textit{balanced-to-unbalanced}) transformer
	used to balance the interferences and the second stage contains a common-mode transformer to filter them.
	The output \SI{5.5}{V} voltage is split into two lines: one (\SI{4}{A} max) is dedicated to the Block Card and it is equipped
	with an over-current protection circuit; the other (\SI{40}{A} max) is used to power the FEE cards and it is controlled
	by a power MOSFET. The \SI{22}{V} output line, together with the \SI{5.5}{V} voltage monitor line,
	is connected to the Power Supply card via the backplane.
	
	The PS card main task is to produce and monitor the voltages required by the Block Card and the FEE cards.
	In particular, it uses DC/DC switching converters to transform the \SI{22}{V} input to \SI{+13}{V} DC (\SI{2}{A}), \SI{-9}{V} DC (\SI{1}{A}),
	\SI{+6}{V} DC (\SI{5}{A}), and \SI{-6}{V} DC (\SI{2}{A}). The \SI{+13}{V} and \SI{-9}{V} lines, which are used to bias the pre-amplifiers,
	have an additional linear regulator to reduce the high frequency ripple. The PS card produces also a negative HV line, currently
	used to bias the CsI(Tl) crystal photodiodes, which is user settable in the range \SIrange{0}{600}{V} DC (\SI{4}{mA} max).
	Due to the huge range of the output voltage, it was not possible to use only a flyback converter. In particular,
	a step down power supply stage (controlled by the same driver) is placed before the standard flyback circuit implemented with the UC3842 driver.
	In this way, we obtained that the voltage at the input of the flyback stage is reduced when the output voltage decreases.
	Moreover, to achieve a better commutation precision of the driver \textit{Pulse-Width Modulation} (PWM) stage, a circuit that modulates the switching frequency
	in function of the set voltage was implemented.
	
	The PS card is also equipped with a PIC microcontroller which continuously measures the voltages and the current flowing through the power supply lines;
	it also monitors the temperature of the board and, in the case of overheating, it shuts down the FEE card lines.
	
	\subsection{FEE cards}\label{ssc:fee}
\begin{figure}[htbp]
		\centering
		\includegraphics[width=\textwidth]{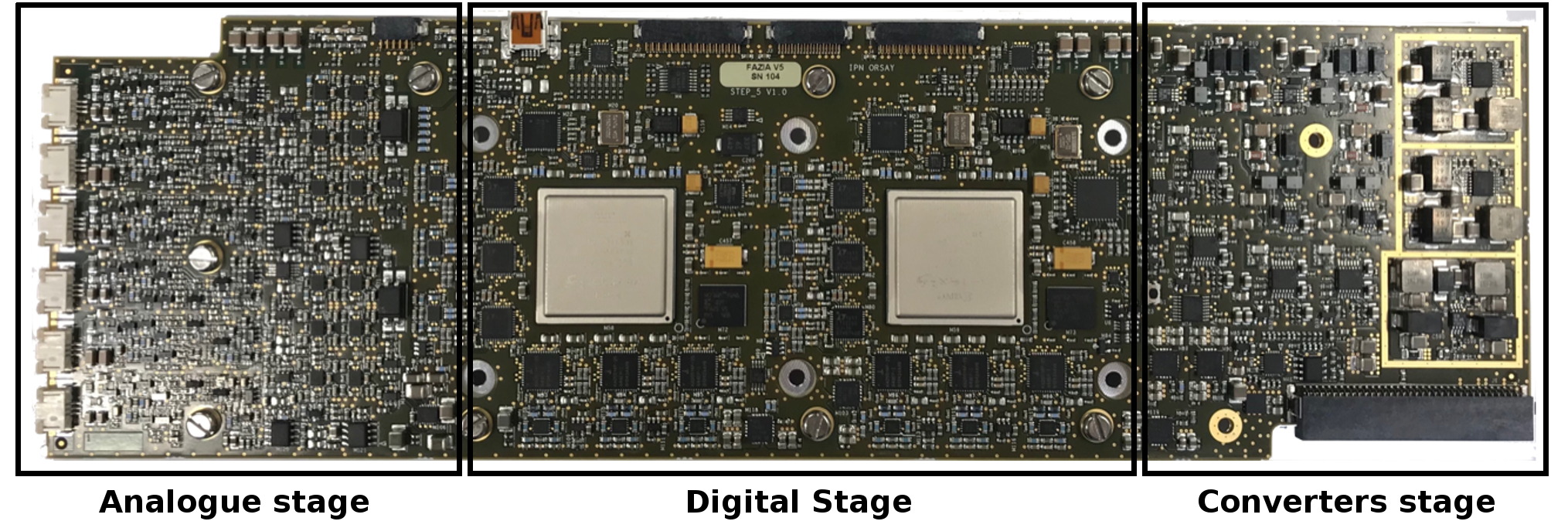}
		\caption{Picture of a front-end electronic card without the protection and dissipation plates.
		The three stages in which the card is subdivided are highlighted.}
		\label{fig:fee}
	\end{figure}

	The core of the FAZIA block is the front-end electronic card \cite{Salomon16}. Up to 8 FEEs can be mounted on each block.
	The area of these boards (\SI{299x88}{mm}) is subdivided into three parts (Fig.~\ref{fig:fee}).
	The first stage embeds all the low-noise analogue electronics (pre-amplifiers, amplifiers and anti-aliasing filters).
	At the opposite side, the FEE hosts the switching power-supplies which are sources of electromagnetic disturbances and
	thus are placed, by design, far from the analogue stages. In the middle part, among many other components,
	two Xilinx Virtex-5 (model XC5VLX50) FPGA chips (one for each telescope and called ``A'' and ``B'' from now on),
	a PIC microcontroller, and 12 ADCs are mounted.
	The printed circuit board has 16 layers on which 1700 components are located on the top surface only.
	The power consumption of a card is about \SI{30}{W}. The connection toward the heat sink is done by
	two aluminium plates: one pressed and sticked on the upper side of both FPGAs and one entirely covering the back side of the card.
	This second sheet has a shelf that is screwed on the main cooled copper plate of the block (see Sec.~\ref{ssc:cooling}).
	
	\paragraph{Analogue stage}
	Six charge pre-amplifiers (three per telescope) are placed on each front-end card just next to the detector connectors.
	Their architecture is based on a folded cascode amplifier and the output dynamic range is \SI{8}{V} for
	a total energy of \SI{4}{GeV} ($\sim\SI{300}{MeV}$ Si-equivalent for the CsI(Tl) channels),
	providing a sensitivity of \SI{2}{mV/MeV} for both Si1 and Si2.
	The detector signals are AC coupled with a \SI{10}{nF} capacitor to block the bias voltage.
	A circuitry (containing a DAC) is added to the pre-amplifier output stage to tune the output offset voltage.
	This function allows one to set the analogue chain baseline close to the bottom level of the ADC input range,
	in order to better exploit the available dynamic range of the ADC.
	The baseline level could be remotely controlled in any moment via a slow control command, since the
	DAC input buses are directly connected to the FPGAs.
	The analogue lines then split to have multiple channels per detector. In particular, for the
	first Si stage, we have three paths: high range charge signal (QH1), low range charge signal (QL1) and current signal (I1).
	For the second Si stage we have the high range charge signal (Q2) and the current signal (I2). For the CsI stage
	we have only one charge signal (Q3).
	The high range (low gain) signals (QH1, Q2 and Q3) are attenuated by a factor 4, to adapt the \SI{8}{V} dynamic
	range of the pre-amplifiers to the \SI{2}{V} input range of the ADCs.
	On the contrary, the low range (high gain) signal (QL1) is amplified by a factor 3. The current signals (I1 and I2)
	are obtained by analogue differentiation of the Si1 and Si2 pre-amplifier outputs.
	All the six signals described above pass through an anti-aliasing filter before being sampled by the ADCs.
	
	A square pulse generator has been built on the front-end card in order to test the analogue chain response.
	It is also useful to verify possible amplification changes during data taking (which, incidentally, have never been observed up to now).
	The generator is based on a simple MOSFET clocked switch connected to the inputs of all the
	pre-amplifiers through a capacitor.
	The user can change the amplitude, the frequency, and the duty cycle of the pulse generator
	via slow control instructions. The pulser amplitude is set by a DAC, whose input bus is connected to
	the front-end PIC microcontroller. Frequency and duty cycle depend on the switch clock, that is generated
	by the FPGA ``B''. Alternatively, an external clock could be used by setting the appropriate slow control register.
	
	\paragraph{Digital stage}
	In the previous paragraph, the twelve signals (six per telescope) that are generated in the analogue stage of the card were described.
	QH1, Q2 and Q3 signals are connected to 14-bit, \SI{100}{MS/s} analog-to-digital converters with a \SI{2}{V} input range.
	Since we have a \SI{2}{mV/MeV} sensitivity from the pre-amplifier and then we reduce the signal by a factor 4,
	at the end we get an energy conversion factor of about 4.1 ADC units per MeV for QH1 and Q2 signals.
	The relatively slow sampling rate is more than sufficient for energy measurements.
	Moreover, the very high effective number of bits ($ENOB=11.4$) of the ADCs guarantees an accurate reconstruction of the released
	energy inside the detector.
	QL1, I1 and I2 signals are sampled by 14-bit, \SI{250}{MS/s} ADCs with a \SI{1.5}{V} input range.
	Considering the QL1 signal, we have again a \SI{2}{mV/MeV} sensitivity from the pre-amplifier but in this case we amplify
	the signal by a factor 3, so at the end we get an energy conversion factor of about 66 ADC units per MeV
	(16 times larger with respect to the corresponding high range signal).
	The low range line is thus very useful to identify and measure energy of particles that produce small energy losses
	within the first Si layer, such as light fragments up to $Z\sim10$.
	Another important feature of the QL1 signal is the possibility to use it for timing, because
	it is acquired by a fast sampling ADC.
	Current signals are also acquired by \SI{250}{MS/s} 14-bit ADCs.
	In fact, because of their fast time evolution, their shape cannot be faithfully reconstructed by interpolation if sampled at \SI{100}{MS/s}.
	Indeed, the current signal represents an accurate image
	of the charge collection process within the detector and allows the best identification
	performances via pulse-shape discrimination for ions stopping in the first stage of the telescope \cite{Carboni12,LeNeindre13,Pastore17}.
	
	The twelve analog-to-digital converters are read by the two FPGAs: the six signals from the first telescope
	are handled by FPGA ``A'' and the six signals from the second telescope are handled by FPGA ``B''.
	The two programmable arrays compute in real time the energy through digital filtering.
	They also generate local triggers, data packing and transmission to the acquisition (see Sec.~\ref{ssc:data}).
	
	Also the 8-bit PIC microcontroller lies in the digital part of the FEE card.
	It is capable to accept and execute commands received through a serial slow control link and
	it can subsequently write and read the slow control FPGA registers by Serial Peripheral Interface (SPI) link.
	As previously said, the PIC controls the pulser amplitude.
	It controls also the high voltage devices (described in the next paragraph) and reads, by means of a dedicated 16-bit ADC,
	the current flowing through the Si detectors.
	When this reverse current increases, the PIC automatically increases the HV device output to compensate
	the voltage drop on the \SI{10}{M\ohm} bias resistor and to maintain the biasing voltage of the silicon detectors at constant value.
	This function is quite important to preserve good PSD during an experiment.
	Finally, the PIC also gets temperature values coming from embedded sensors located in several critical positions
	on the cards.
	
	\paragraph{Converters stage}
	The voltage conversion stage is placed just close to the backplane connector.
	The low voltage power supply part includes four switching converters and more than 20 linear
	regulators. Four high voltage devices are also embedded on the board. The high voltages
	are generated from a common \SI{5.5}{V} input and their purpose is to bias the silicon detectors
	of the two telescopes. The architecture of the HV modules is based on a switching regulator and a transformer.
	The high voltage devices for Si1 can ramp up to \SI{300}{V} with a precision of less than \SI{0.1}{V}.
	The HV device for Si2 can ramp up to \SI{500}{V} with the same absolute precision.
	The HV functions and values are remotely controlled via the PIC microcontroller.
	The integrated high voltage device represents a new and original solution within the nuclear physics
	community, as it is embedded into one single electronic card operating under vacuum.
	
	\subsection{Block card}\label{ssc:bc}
\begin{figure}[htbp]
		\centering
		\includegraphics[width=0.8\textwidth]{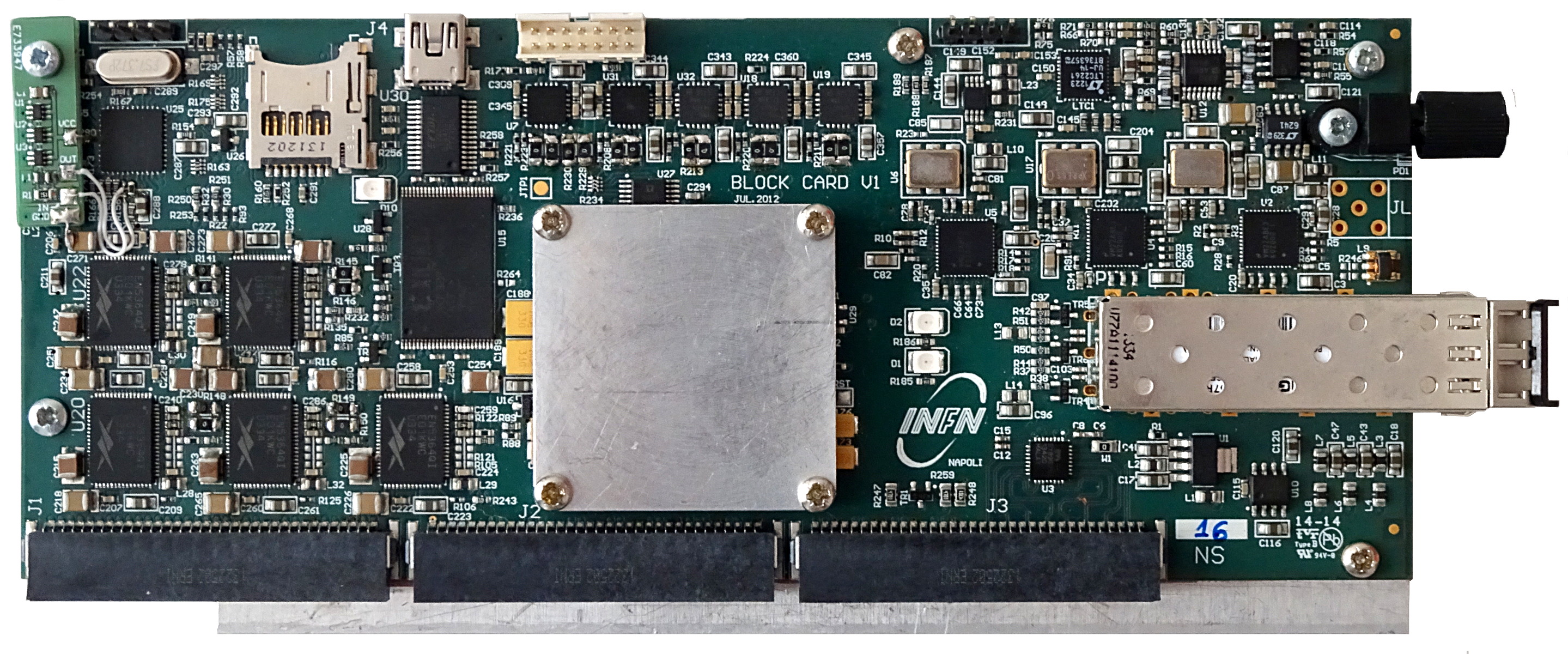}
		\caption{
			Picture of a Block Card. In the center, the FPGA is covered by the heat sink. In the right side, the SFP transceiver
			used to communicate with the Regional Board can be seen.
		}
		\label{fig:blkcard}
	\end{figure}

	The main task of the Block Card (shown in Fig.~\ref{fig:blkcard}) is to retrieve all the data coming from the FEE cards and to build from them a
	partial event. The main communication path between the BC and FEE cards is implemented with 24 serial
	buses in a ``star'' configuration: each front-end card is reached by three \SI{400}{Mbit/s} buses.
	Two buses connect each FEE card to the BC and one goes from the BC to each FEE.
	The implementation of these fast serial links was done using Xilinx ``ISERDES'' and ``OSERDES'' cores.
	To align and synchronize data, a programmable delay was also added before the ISERDES unit.
	Moreover, other common signals reach every FEE card from the Block Card: i.e. the validation signal,
	the external pulser clock and the slow control line. Every kind of communication between
	BC and FEEs passes through the backplane. On the contrary, the communication with the event building electronics
	takes place through a \SI{3}{Gbit/s} optical link. In particular, the Block Card hosts a \textit{small form-factor pluggable} (SFP)
	optical transceiver that supports two (RX and TX) SX fibers (\SI{850}{nm} wavelength) with a LC connector.
	
	Communications and data sorting are handled by a Xilinx Virtex-5 FPGA (model XC5VFX70T), together with
	the fundamental task of capturing a special packet from the optical link to produce a synchronized clock (see Sec.~\ref{ssc:clock}).
	The Block Card contains also a photodiode which reads another optical fiber and a \SI{50}{MS/s} ADC to sample
	the signal coming from it. To exploit the full ADC range, a variable gain amplification circuit was also implemented.
	This system is used as an extra synchronization method, as it will be described in Sec.~\ref{ssc:clock}.
	Finally, the BC features a microSD slot. If a memory card with a special file is present in the slot when the board is switched on,
	the block identification number (ID) is read from that file. In this way it's very easy to change the block ID during the experiment.
	The block ID is the only way to distinguish the blocks when a slow control command is sent. The microSD card could also be
	used to reload the FPGA firmware with a ``.xsvf'' file.
	
	\subsection{Event building electronics}\label{ssc:rb}
\begin{figure}[htbp]
		\centering
		\includegraphics[width=0.6\textwidth]{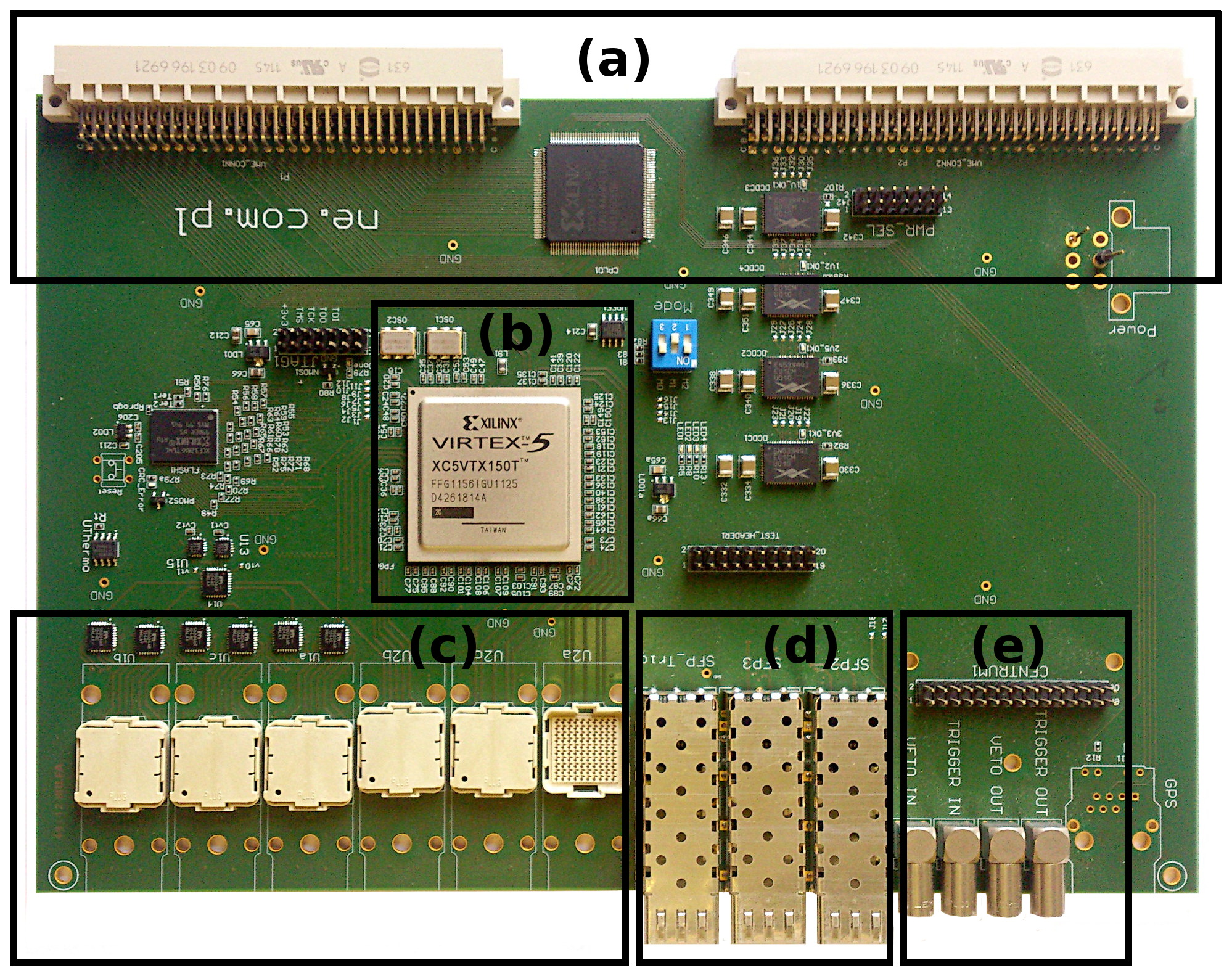}
		\caption{Picture of the regional board with highlights of the most important features: (a) VME connectors and CPLD,
		(b) FPGA and crystal oscillators, (c) special connectors for block communication; (d) SFP slots; (e) LEMO connectors and CENTRUM port.}
		\label{fig:rb}
	\end{figure}

	The regional board (Fig.~\ref{fig:rb}) is a 6U size VME card (operating outside the reaction chamber, in air) whose main tasks are: to read data from the FAZIA blocks
	and form a complete event, to analyze the triggers from the blocks and possibly send validations to them,
	to handle slow control requests from many PCs and eventually to communicate with acquisition servers.
	The RB is an evolution of the ``test card'' developed by INFN -- Naples group. That card featured an USB protocol to transmit data to the acquisition system and it was capable
	to handle up to 8 blocks. The test card was widely used to develop many features implemented in the regional board, such as the internet protocol on the on-board FPGA.
	
	The RB can manage up to 36 blocks and, in case of configurations that need a higher number of blocks,
	multiple regional boards may be interconnected using optical links and/or the VME bus. However, this feature
	has not been yet implemented.
	
	In addition to the VME bus, regional board connections include six special connectors for block communications,
	three SFP slots for RB interconnection and communication with acquisition (through Ethernet protocol),
	four LEMO connectors (``veto in'', ``veto out'', ``trigger in'' and ``trigger out'') in LVTTL logic
	for trigger coupling with other devices and a port for event synchronization among many compatible detectors.
	The connectors for block communications allow the installation of an optical translator over each of them.
	At the other end of the optical translator, a compact 12-fiber connector is placed. Each translator
	handles up to twelve mono-directional fiber connections (RX or TX), thus two translators are needed to operate up to twelve blocks
	(four translators for up to 24 blocks and all the six translators for up to 36).
	The port for event synchronization uses the CENTRUM technology developed at the GANIL laboratory (Caen, France).
	This technology was chosen in view of the forecast coupling of FAZIA with the INDRA detector array,
	which is installed at GANIL and already uses CENTRUM protocol.
	All the features and communications (except the VME bus management) are handled by a Xilinx Virtex-5 FPGA (model XC5VTX150T).
	The VME bus is operated by a Xilinx CPLD (model XC95144XL) that is directly connected to the FPGA.
	The \SI{5}{V} operating voltage is supplied by the VME bus.
	
	\section{Functional description of the electronics}\label{sec:functional}
	\subsection{Clock distribution}\label{ssc:clock}
\begin{figure}[htbp]
		\centering
		\includegraphics[width=0.6\textwidth]{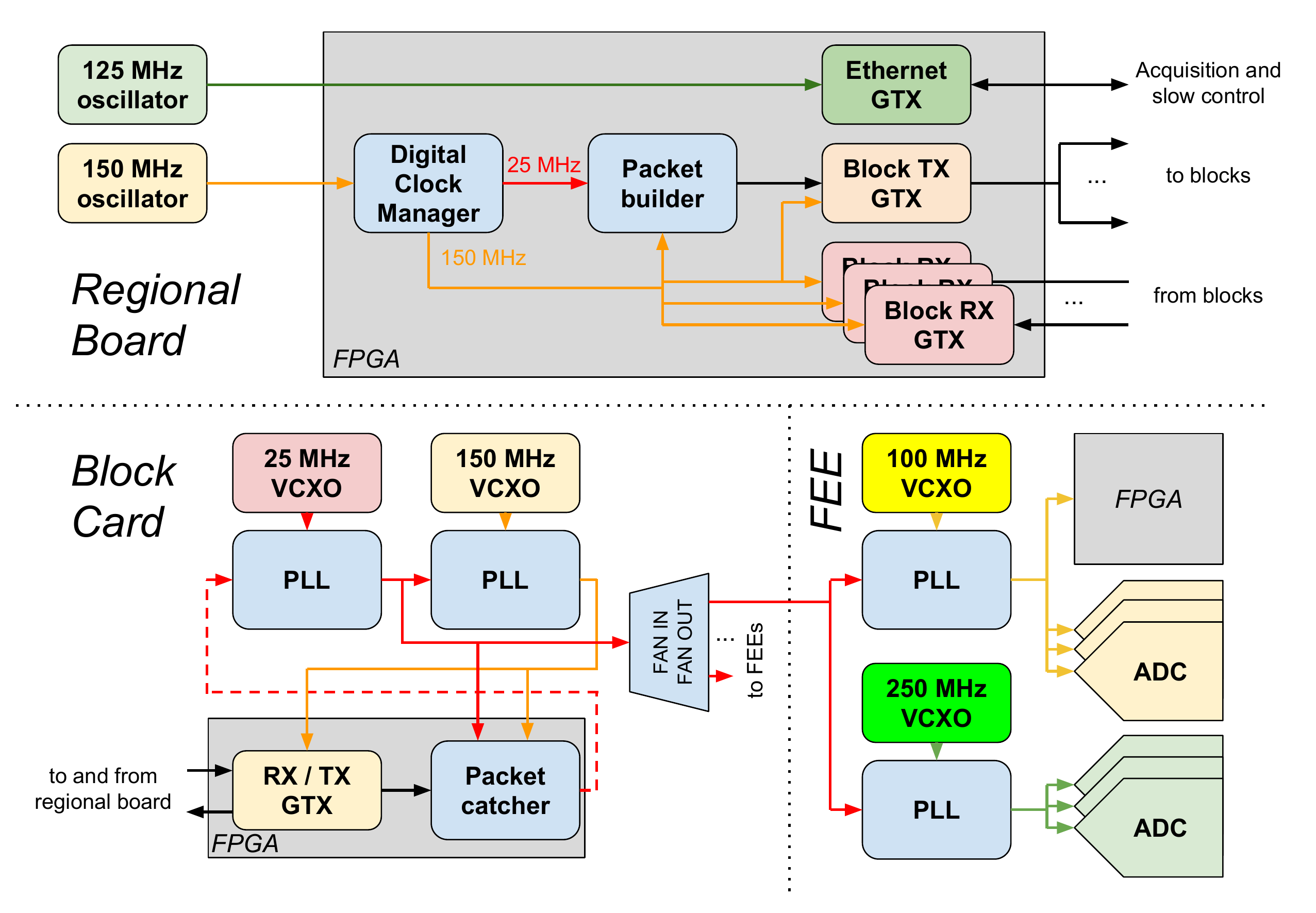}
		\caption{Schematic representation of the FAZIA clock distribution, which was designed in order to sample all the signals with the same phase. Further details in the text.}
		\label{fig:clock}
	\end{figure}

	FAZIA is designed to measure over a broad range of beam energies. On some application, mainly with low energy beams,
	the time of flight (ToF) technique is very useful and the clock synchronization among all the ADCs is necessary.
	In fact, if the ADCs in different blocks had independent clocks,
	the accuracy of the time measurement could not be better than one clock cycle (\SI{4}{ns} or \SI{10}{ns} depending on the kind of ADC).
	So, to synchronize all the sampling ADCs, they must be provided with exactly the same clock on all the cards.
	The clock distribution tree is schematically represented in Fig.~\ref{fig:clock} and it is detailed here.
	The primary clock is generated on the regional board: there are two crystal oscillators
	set at \SI{125}{MHz} and \SI{150}{MHz} and both are connected to the FPGA chip. The former is used only to
	clock the Ethernet part of the FPGA project. The latter is used to generate (inside the FPGA) a \SI{25}{MHz} clock,
	and to clock the built-in GTX transceivers, which are used to send and receive data from the blocks through the
	optical link. The transceivers are devices embedded inside the FPGA. In our case
	they convert 16-bit data at \SI{150}{MHz} (\SI{2.4}{Gbit/s}) to a serial line at \SI{3}{Gbit/s} and vice versa. The missing 1.25 factor
	in the data rate comes from the 8b/10b coding (see Sec.~\ref{ssc:packet}) of the serial line.
	Xilinx GTP and GTX transceivers are very suitable to implement a connection with fixed latency,
	as they can provide an extremely reduced clock skew \cite{Giordano11}.
	
	The optical fibers leaving the regional board enter the scattering chamber and reach the various blocks.
	They are connected to the SFP optical translators on the block cards. The signals eventually arrive inside the
	block card FPGAs, where they are de-serialized by GTX transceivers. As briefly introduced in the previous section,
	the BC has a peculiar system to recover the \SI{25}{MHz} clock from the optical link:
	on the card there is a voltage controlled crystal oscillator (VCXO), connected to a phase-locked loop (PLL) device, which is used to clock the FPGA.
	The PLL reference is the signal recovered from the fiber by the FPGA itself. So, when the BC is switched on,
	the PLL is not locked and the FPGA is clocked by a \SI{25}{MHz} signal that is uncorrelated with the
	\SI{25}{MHz} signal generated on the regional board. Then the block card starts to catch the special K28.5 sequences (see Sec.~\ref{ssc:packet})
	from the optical link and generates the synchronized \SI{25}{MHz} signal that enters into the PLL device.
	At this point the PLL is locked and so is the FPGA clock. The PLL output is also split into eight outputs
	that reach the FEE cards through the backplane. The front-end cards use the phase-locked
	\SI{25}{MHz} signal to produce (using VCXO and PLL devices) \SI{100}{MHz} and \SI{250}{MHz} frequencies.
	Finally, these signals are used to clock the ADCs and the FPGAs. Due to the physiological delays between clock edges and ADC sampling,
	practically the various sampled signals are not exactly synchronous, since the delays are not identical among the ADCs. The methods to reduce
	the residual asynchronism, of the order of \SI{100}{ps}, are currently under study and they will be the subject of another work.
	
	The phase-locked \SI{25}{MHz} clock on the FEEs is used also to increment a universal 15-bit timestamp counter.
	To ensure that all the cards are assigning the same timestamp at the same moment, a ``time tag zero'' signal (TTSync)
	is sent on the optical fibers by the RB every time that the timestamp counter goes overflow (\SI{1.31}{ms}).
	
	To improve the block cross-synchronization, an external sinusoidal signal (the same for all blocks)
	could be sampled by the block cards at \SI{50}{MS/s}. Time marks extracted from sinusoidal waveforms
	are very precise, even when the sampling rate is not very high \cite{Bardelli07}. This method
	allows to check the clock synchronization among blocks with a precision down to \SI{10}{ps}.
	The \SI{50}{MHz} clock is generated by means of VCXO and PLL from the synchronized \SI{25}{MHz} clock,
	as also the \SI{150}{MHz} reference used to clock the GTX devices on the Block Card.
	
	\subsection{Packet structure}\label{ssc:packet}
	Since each block is connected with the regional board using only an optical link, the packet structure
	should be optimized to reduce the data overhead while transmitting all the needed information.
	The optical link is composed of two fibers: one is used to transmit data to the block (TX) and the other
	to receive (RX). In both directions data are structured in packets of six 16-bit words and the 8b/10b encoding is used.
	That means that each byte, when is sent on the fiber, is converted into a 10-bit frame: this conversion helps
	to provide enough state changes to allow clock recovery. Converting 8-bit sequences to 10 bits means also
	that some 10-bit sequences have no corresponding 8-bit data. One of these sequences (called ``K28.5'') is
	used to distribute and synchronize the master clock among all the blocks (see Sec.~\ref{ssc:clock}).
	\begin{table}[htbp]
		\centering
		\begin{center}
			\begin{tabular}{ccc}
			\toprule
			\textbf{Word} & \textbf{TX} & \textbf{RX}\\
			\midrule
			0 & sync and control & sync and control\\
			1 & event number    & data\\
			2 & -                & data\\
			3 & block acq. busy  & data\\
			4 & block acq. busy  & data\\
			5 & block acq. busy  & data\\
			\bottomrule
			\end{tabular}
		\end{center}

		\caption{Summary of the packet structure used in the data transmission between the RB and the blocks. More details in the text.}
		\label{tab:packet}
	\end{table}
	
	The first 16-bit word of the packet is always a special synchronization and control frame, both in TX and RX transmission.
	In particular, the first 8 bits are not defined and they are substituted by the K28.5 sequence in the 8b/10b conversion.
	The whole packet structure is summarized in Tab.~\ref{tab:packet} and detailed in the next two paragraphs.
	
	\subsubsection{TX packet structure}\label{sss:tx}
	As said before, the first frame of the transmitted packet is used by the block card to synchronize the master clock via the K28.5 sequence.
	The remaining 8 bits have the following use:
	\begin{description}
	 \item[bit 7] validation signal generated by the RB (see Sec.~\ref{ssc:trigger});
	 \item[bit 6] slow control TX line (see Sec.~\ref{ssc:slow});
	 \item[bit 5] TTSync signal (see Sec.~\ref{ssc:clock});
	 \item[bit 4] global reset of all blocks;
	 \item[bit 3:0] not assigned.
	\end{description}
	When a validation signal is sent, the second frame of the TX packet contains the 12-bit event number used to check the consistency
	in the event building phases (see Sec.~\ref{ssc:data}). The third frame is not used. The last three frames are used to selectively block
	the data transmission from the blocks. In fact, when a regional board FIFO dedicated to a specific block is about to be completely full,
	the RB must block the data acquiring process only from that block. That is achieved by writing on a TX frame
	a block ID followed by its acquisition status (enabled/disabled). The involved block always reads the last three TX frames
	searching for its ID: if it is found, the block reads the acquisition enable bit and updates its own status.
	
	\subsubsection{RX packet structure}
	The first frame of the received packet is, also in this case, a synchronization and control word and the first part is again the K28.5 sequence.
	The remaining 8 bits of the first RX frame are the following:
	\begin{description}
	 \item[bit 7] GTT flag (see Sec.~\ref{ssc:data});
	 \item[bit 6] slow control RX line (see Sec.~\ref{ssc:slow});
	 \item[bit 5:1] block trigger multiplicity (see Sec.~\ref{ssc:trigger});
	 \item[bit 0] not assigned.
	\end{description}
	The other five frames of the packet contain the data flow coming from the block.
	
	\subsection{Trigger logic}\label{ssc:trigger}
	On both FPGAs of the front-end cards, fast trapezoidal shaping filters are implemented
	on the QH1, Q2 and Q3 signals in order to generate local triggers.
	There are also slow control registers to adjust, for each channel, the filter parameters (rising edge and flat top lengths),
	the low threshold and the high one. Usually, the filter is set at a \SI{200}{ns} rising edge and a \SI{200}{ns} flat top:
	these values do not affect the sustainable event rate, which is limited by the data acquisition (see Sec.~\ref{ssc:data}).
	For each front-end, the user can also choose the trigger timeout,
	the trigger source (logic OR among any combination of Si1, Si2 and CsI) and the kind of threshold: i.e.
	one may use the low threshold only (trigger is produced when the maximum amplitude of shaped signal is larger than it) or both low and high
	thresholds (trigger is produced when the maximum amplitude of shaped signal is between them).
	The local triggers generated by the FEEs reach the block card through 16 dedicated lines (one per telescope) on the backplane.
	
	FAZIA trigger system is multiplicity based: on each block, the BC counts the local triggers and sends the total
	to the regional board every \SI{40}{ns} through the optical link.
	The RB collects all the multiplicity values coming from each
	block and applies up to eight programmable rules. For each one the user can choose (via slow control) the blocks checked
	by the rule, the multiplicity threshold and the downscale factor $K$. The regional board will then integrate inside a time window
	the multiplicities coming only from the blocks specified by the rule, and it will produce a ``rule trigger''
	only if the integrated value overcomes the multiplicity thresholds. The trigger is then accepted once every $K$ occurrences.
	The logic OR among all the rule triggers is eventually the global trigger signal.
	The RB then checks if there are any alerts: FPGA data buffers are almost full, the GTT flag (see Sec.~\ref{ssc:data}) from any block is true
	or there is an external veto from the ``veto in'' LEMO connector; if there is at least one alert, then a veto flag is issued.
	In these cases, except when there is only an external veto, the flag is also sent to the ``veto out'' LEMO connector.
	The ``trigger out'' LEMO connector is true when there is a global trigger without the veto flag.
	The main output of the trigger component on the regional board is the validation signal: it is
	produced when the veto flag is false and the global trigger or the external trigger from the ``trigger in''
	LEMO connector are true. The user may also choose to work in ``slave'' mode (when coupling FAZIA with other devices):
	in this case the validation is only produced when the veto flag is false and the external trigger is true.
	In any case the validation signal is sent at a \SI{25}{MHz} rate to every block via the optical links,
	captured by the block cards and distributed to every FEE through the backplanes.
	Together with the validation signal, also an event number, generated by a counter on the regional board FPGA,
	is always sent to the FEEs using the same path.
	
	\subsection{Data flow}\label{ssc:data}
	On both FPGAs of the front-end cards, four trapezoidal shaping filters are implemented,
	in addition to the trigger shapers described in the previous section, to calculate in real time the energy
	released in each stage of the telescope. There is one shaper on the QH1 signal, one on Q2 and two on Q3.
	In fact, the CsI(Tl) signal has a ``fast'' and a ``slow'' filter in order to exploit the well known $fast-slow$ technique
	\cite{Gal95} to identify the light ions that stop
	in the last stage of the telescope. As in the case of the trigger shapers, all the rising edge and
	flat top lengths of the filters are user adjustable.
	Usually, the filters on QH1 and Q2 were set at a rise time of \SI{2}{\micro s} and a flat top of \SI{1}{\micro s}.
	Rise time and flat top for Q3 shapers were set respectively at \SI{2}{\micro s} and \SI{10}{\micro s} for the slow filter
	and at \SI{2}{\micro s} and \SI{500}{ns} for the fast one. These values do not affect the sustainable rate because
	it is limited by the maximum acquisition rate ($\sim$\SI{2000}{ev/s}). Since a faster rate is not affordable, the beam intensity can be accordingly chosen to avoid pile-up and dead time.
	In addition to being shaped, all the signals (including also QL1, I1 and I2 that have no shaper) are continuously stored in circular buffers with dimension $N\leq 1024$, settable via slow control.
	
	When a front-end card FPGA receives a validation, all the raw signals are transferred from their circular
	buffers to FIFO memories, whose lengths are adjustable via slow control up to a maximum of 4096
	samples ($\sim\SI{41}{\micro s}$) for QH1, Q2 and Q3 signals and 8192 samples ($\sim\SI{33}{\micro s}$) for QL1, I1 and I2.
	At the same time, the acquisition thresholds are checked.
	These thresholds are again user programmable via slow control and act on the energy shaped signals.
	If any of the QH1, Q2, Q3 ``fast'' and Q3 ``slow'' shaped signals exceeds its respective acquisition threshold,
	then all the telescope signals are marked for acquisition.
	A single large FIFO memory is finally used to store the whole local event from the telescope handled by the FPGA.
	On this memory, if the telescope is marked for acquisition,
	the six waveforms are transferred together with the four maximum values of the shaped signals and
	the event number sent by the RB. Between each front-end card and the block card,
	the data travel on two \SI{400}{MHz} serial buses (introduced in Sec.~\ref{ssc:bc}),
	thus capable to offer a throughput of \SI{800}{Mbit/s}. The buses are connected to the FPGA ``A'' on the FEE side,
	so data sampled by the telescope ``B'' must pass through FPGA ``A''. If any of the FIFO memories on the front-end card
	is about to fill up, the \textit{Global Trigger Throttle} (GTT) flag is raised and the RB is vetoed.
	
	When a block card FPGA receives the validation signal, it enters a state where it starts to read the FIFO memories
	on the front-end cards.
	The event number written inside the data coming from each FEE is checked and the event is discarded if the number is
	less than expected. If the actual number is greater than expected, instead, the FEE is skipped
	but the data are kept for the next event. In this way the block card builds a coherent partial event and stores it in
	a FIFO buffer waiting to transfer it to the regional board. Inside the partial event, also the external sinusoidal
	signal (see Sec.~\ref{ssc:clock}) is stored.
	
	After the generation of the validation signal, the regional board enters a state where it starts to read the FIFO memories
	on the block cards.
	In complete analogy to the block card behavior, the RB checks the event number written inside the data coming
	from each block and the event is discarded if the number is
	less than expected. If the actual number is greater than expected, instead, the block is skipped
	but the data are kept for the next event. In this way the regional board builds a full coherent event and stores it in
	a large FIFO buffer. The RB adds also some trigger information (i.e. number of accepted and vetoed triggers in the last ten seconds)
	to allow the calculation of the dead time.
	The FIFO memory is continuously read by a component of the FPGA code that produces Ethernet frames
	using the \textit{User Datagram Protocol} (UDP) and send them to the acquisition system through a optical fiber connected to a SFP translator.
	The user may specify up to 16 machines (sending their IP and MAC addresses via slow control) to which data will be sent.
	The RB will send an event to the first computer, then another to second and so on. Then it will start again from the first.
	In this way the dead time due to computing time is minimized. The maximum throughput obtained with the Ethernet
	communication is about \SI{800}{Mbit/s}.
	
	\subsection{Coupling with other detectors}\label{ssc:coupling}
	The regional board features some programmable auxiliary connections which can be used to couple FAZIA with
	other detectors. In particular, they will be used soon to measure together with INDRA at GANIL.
	The coupling is done on different levels. The lowest is the trigger level: to ensure
	a common dead time between FAZIA and INDRA we need to properly interconnect the trigger in/out and veto in/out
	LEMO connections (see Sec.~\ref{ssc:trigger}) between the two detectors.
	The second level consists in the generation of a timestamp: this is performed thanks to the CENTRUM module,
	which is an absolute timestamp generator connected to both apparatuses.
	When FAZIA generates a validation signal, it also sends a timestamp request to the CENTRUM system, which dispatches to the RB a frame containing a 48-bit timestamp and a 32-bit event number,
	plus a 16-bit checksum. A similar packet is sent to INDRA when it produces a request.
	The FAZIA regional board, after checking the checksum, inserts the CENTRUM frame inside the data flow that is sent to the acquisition.
	The third and highest level is done by NARVAL, an acquisition system developed by IPN Orsay (France). NARVAL receives
	data from both INDRA and FAZIA acquisitions and merges those events with timestamp differences smaller than
	a pre-defined (and reaction-dependent) coincidence window, producing global events composed of data from both the apparatuses.
	
	Of course this system is designed to be as general as possible and it may be used in the future also to couple
	FAZIA with other detectors than INDRA. Moreover, both CENTRUM and NARVAL technologies support the connections
	of many apparatuses, so one may also think to easily couple three or four different devices with FAZIA.
	
	\subsection{Acquisition}\label{ssc:acq}
	The FAZIA acquisition (DAQ) has been developed and it is currently maintained by INFN -- Naples. 
	FAZIA DAQ is a multi-threaded and multi-machine system, written in C++ language and consisting of different classes (DAQ modules) that exchange
	messages and events through ZeroMQ sockets \cite{zmq}. The main DAQ modules are the following:
	
	\begin{description}
		\item[FzReader] It acquires raw data coming from the Regional Board by listening to the dedicated UDP socket. Then it forwards the data to FzParser thread pool.
		\item[FzParser] Each FzParser includes a \textit{Finite State Machine} (FSM) able to analyze and validate each acquired event
		in order to put all the information inside a structured format based on the Google Protocol Buffers \cite{protobuf}.
		Multiple FzParser threads can run on a multi-core machine in order to benefit from parallel execution of tasks.
		Each thread eventually forwards data to the FzWriter module through a \SI{10}{Gbit/s} dedicated network.
		\item[FzWriter] This module stores data in files and directories with Google Protobuf data format. 
		It also runs a data spy in order to allow on-line data processing and analysis by external data visualization tools.
		\item[FzNodeManager] It is a local supervisor for FzReader/FzParser or FzWriter that run on each FAZIA DAQ deployed machine. 
		It sends a report on module status to FzController and it receives run control and setup commands for module management.
		\item[FzController] It is a global supervisor for all FzNodeManager modules. 
		It offers a global view on whole cluster status and it accepts commands for FAZIA DAQ setup and run control.
	\end{description}
	
	\begin{figure}[htbp]
		\centering
		\includegraphics[width=0.5\textwidth]{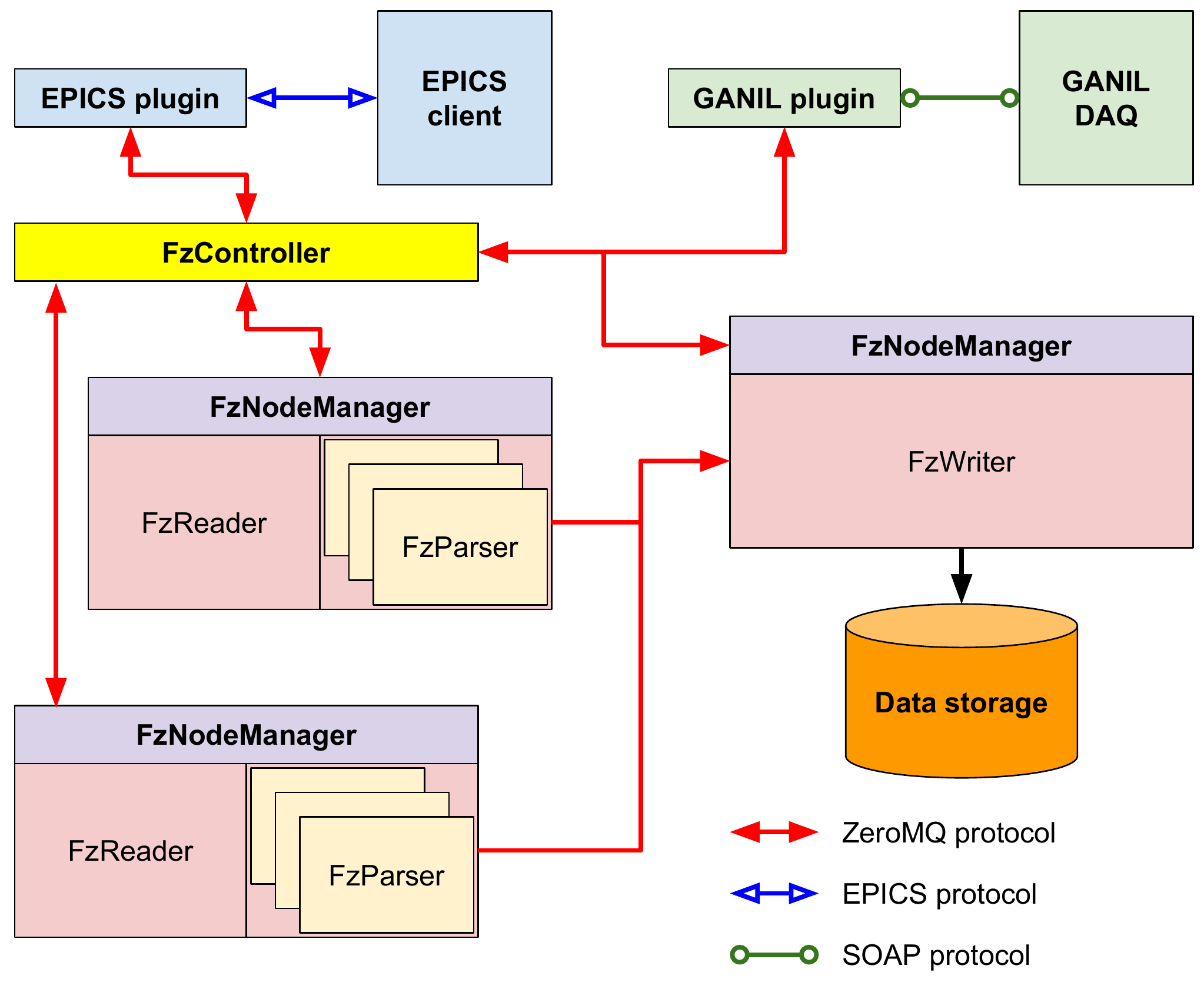}
		\caption{Schematic representation of the FAZIA DAQ system}
		\label{fig:daq}
	\end{figure}
	Fig.~\ref{fig:daq} shows the overall architecture of FAZIA DAQ. Some plugins have also been developed to interact
	with different Run Control systems. These plugins profit of the ZeroMQ network layer of FzController in order to
	control the data acquisition from different clients (e.g. EPICS or SOAP). This feature of FAZIA DAQ makes it
	very suitable and flexible for integration and coupling with other experiments and detectors.
	For example, a GANIL plugin was developed to couple FAZIA with INDRA: it is a C++ class of the DAQ which allows the remote control of FAZIA
	acquisition from NARVAL system using the SOAP protocol. At the same time, the plugin sends data to NARVAL using a TCP/IP connection.
	
	\subsection{Slow control}\label{ssc:slow}
	As illustrated in the previous sections, slow control instructions permit to control almost every aspect of the electronics.
	The commands, in the form of UDP packets, can be sent by any PC in the same subnet where the regional board is located. The RB analyzes the frame
	to check if the instruction is for the regional board itself. If it is, then the board executes the command and immediately sends a reply.
	If not, the RB forwards to all the blocks the message via the optical links. On every block, the instruction is dispatched via the backplane
	to every card containing a PIC (FEEs, PS, and BC). Since the slow control frame uniquely identifies the card
	to which it is intended, only that card answers and the reply is returned to regional board. In every case, the RB transmits back the reply message
	to the PC that has sent the request.
	
	When the slow control request is not for the RB, it must be converted into a standard \SI{115.2}{kbit/s} serial signal which can be correctly read by the PIC devices
	on the PS, BC or the FEEs. In particular, in the regional board FPGA code, a \textit{universal asynchronous receiver-transmitter} (UART) device is implemented.
	The UART converts the slow control frames received via the Ethernet device into a slow serial data flow which is sent on the optical fibers
	to all the blocks. Of course, since the slow control bit is sent on the fiber at a \SI{25}{MHz} rate (see Sec.~\ref{sss:tx}), the \SI{115.2}{kbit/s} serial data flow is oversampled.
	Vice versa, when the RB receives a slow control reply from a block as a serial data flow, this flow is deserialized by the same UART component described above.
	
	\section{Mechanical solutions}\label{sec:mechanics}
	\subsection{Detector holding mechanics}\label{ssc:block}
\begin{figure}[htbp]
		\centering
		\includegraphics[width=0.6\textwidth]{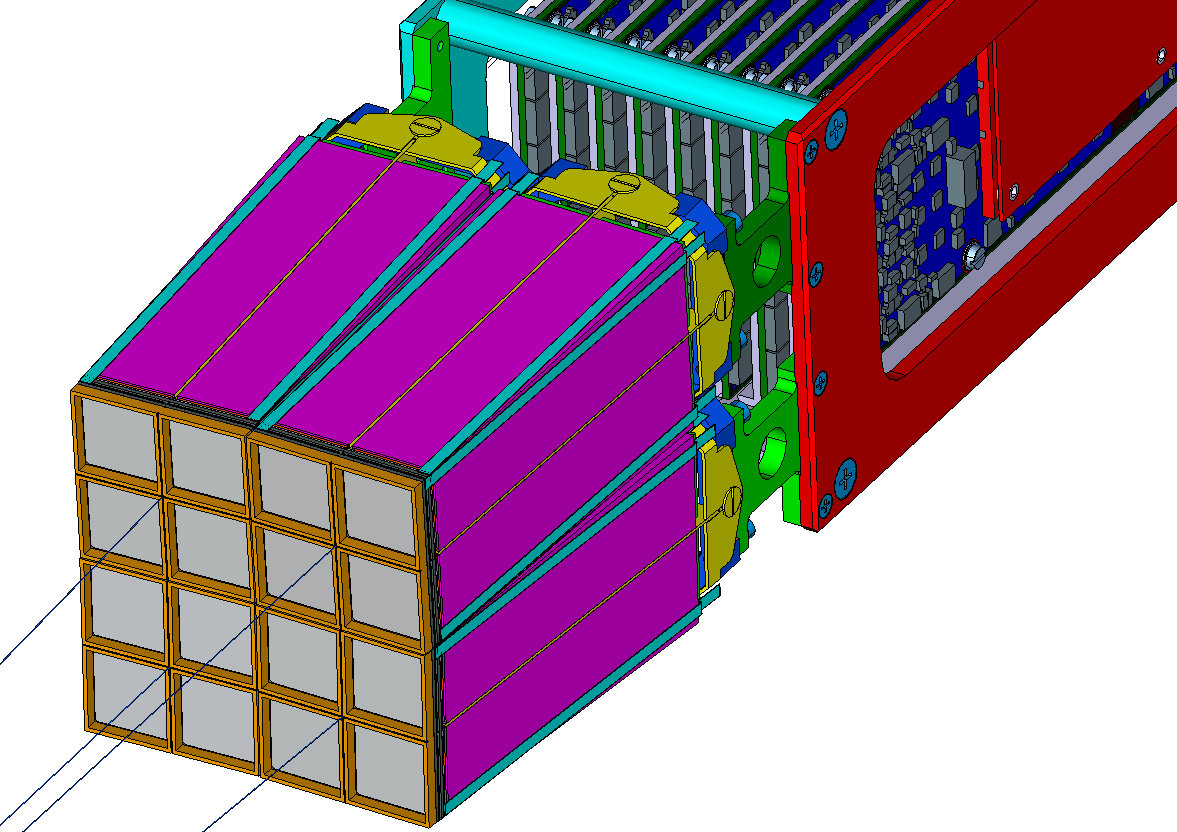}
		\caption{
			Rendering of the detectors and their support mechanics. Orientation axes of quartettos are also shown.
		}
		\label{fig:det}
	\end{figure}

	A rendering of the FAZIA detectors is shown in Fig.~\ref{fig:det}, where silicon pads and CsI(Tl) crystals can be spotted.
	The 16 telescopes, which form a block, are arranged in a $2\times 2$ matrix of \textit{quartettos}.
	A ``quartetto'' is a self-consistent sub-structure of four telescopes in a $2\times 2$ configuration.
	In the figure, the pad precise alignment is also shown. Indeed, each quartetto axis (perpendicular to the surface of the four Si pads)
	points to the target. This configuration has been achieved through a fixed geometry and thus without degrees of freedom.
	Our choice has been to build the various supports for a distance of \SI{100}{cm} between the target and the first Si layer.
	This value has been chosen as a compromise between detector granularity and solid angle coverage. Moreover,
	\SI{100}{cm} is the minimum distance from the target which guarantees a negligible channeling contribution (see below) and
	a sufficient flight base for time of flight identification.
	
\begin{figure}[htbp]
		\centering
		\includegraphics[width=0.5\textwidth]{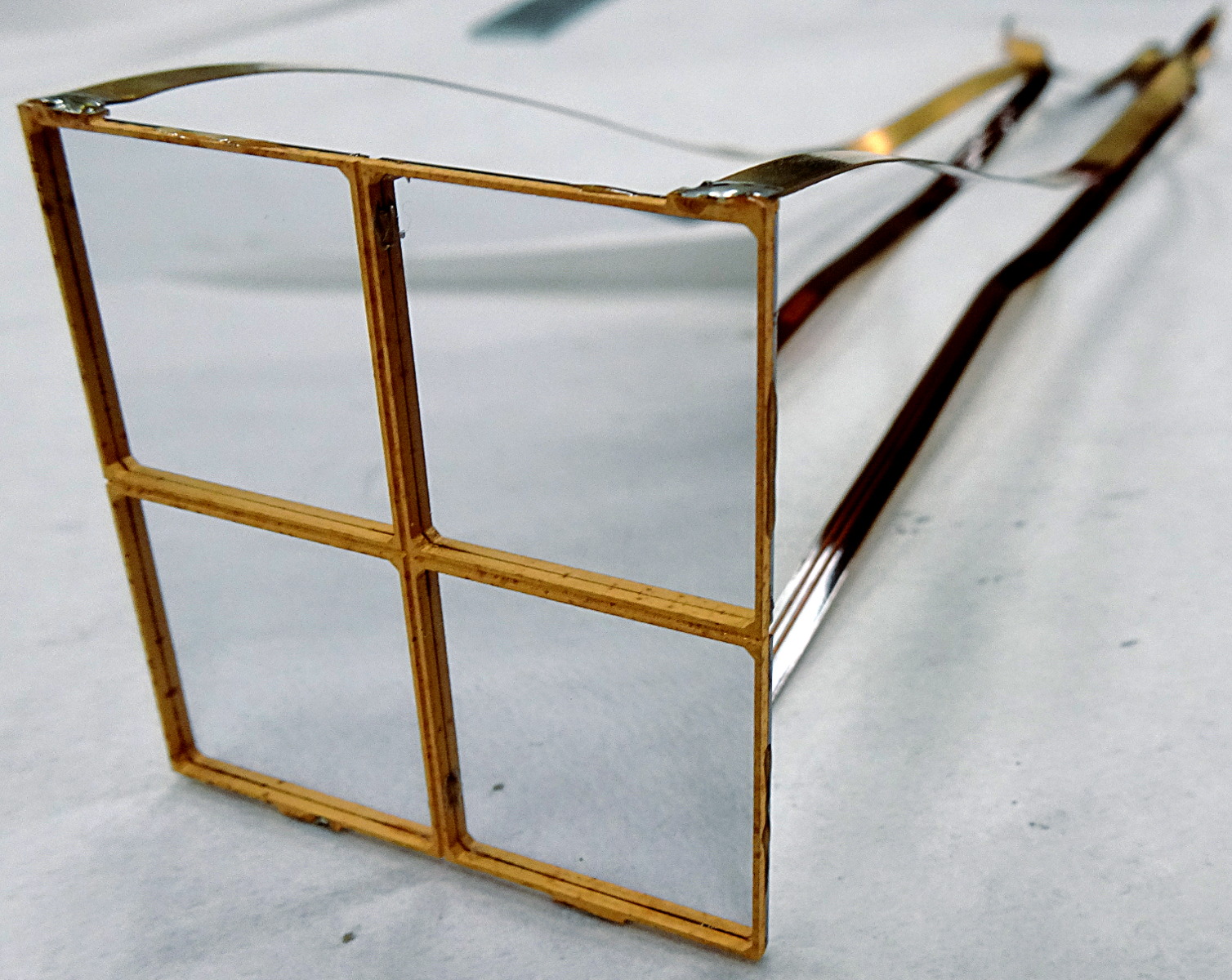}
		\caption{
			A silicon pad holder with mounted detectors is shown: the grooves and the ridges needed to lock two holders in position could be seen. The kapton
			strips, used to connect the sensors to the front-end electronics, are soldered to the metal and $\mu$-bonded to Si pads.
		}
		\label{fig:telaio}
	\end{figure}

	The detector supporting mechanics has been designed focusing on two main goals. The first is the reduction of all the dead
	regions between quartettos and on external edges. As a matter of fact, considering a single block,
	the active area is \SI{84}{\%} of the total front side. The second important goal is the precision of the telescope
	orientation, in order to reduce the channeling effects. In fact, our silicon detectors are cut in such a way that the particles which perpendicularly impinge
	on the sensors do not travel along crystal axes \cite{Bardelli09,Bardelli11}.
	By means of a laser mirroring method (see Sec.~\ref{ssc:laser}) applied to the reflective Si pad surfaces, we obtained that the orientation misalignment is typically less
	than \SI{1.5}{\degree}; larger values have been rarely observed in pads with misplacements or bad gluing. Those two objectives imposed the
	choice of a proper material to build silicon pad holders. We chose the 7075 aluminium alloy, also used in avionics,
	because it is precisely machinable and light but, at the same time, it is robust even in thin layers.
	The supports for the $2\times 2$ Si pad matrices (Fig.~\ref{fig:telaio}) were built using \textit{wire electrical discharge machining}.
	This technique, which softly acts on the bulk, reduces the strains so that the final piece maintains its planarity and details, needed for a good and
	precise gluing of the silicon pads and the matching between different parts.
	We remind that a telescope is composed of two silicon detectors followed by a CsI(Tl) crystal so, to form a quartetto,
	two similar $2\times 2$ silicon frames should be locked one on top of the other by means of very thin grooves and ridges.
	The four CsI(Tl) scintillators are fixed to a central cross-shaped support, which is again made of 7075 aluminium alloy. The strength of this alloy is
	particularly important for this piece, since it must hold the four crystals (weighing \SI{0.72}{kg}) and all the silicon pads as a cantilever.
	
	The fastening between the detectors and the rest of the block is relatively simple, in order to
	allow an easy replacement of the detector ``nose''. In this way, when some detectors are 
	damaged, one can replace the whole nose in a reasonable time. Usually, the insertion of the kapton strips
	into the FEE female connectors is the longest operation.
	
	The above described design and its practical implementation are particularly delicate and many 
	technicians and engineers, from Bologna, Florence and Naples INFN departments, contributed with their work and expertise.
	
	\subsection{Frame for the INDRA-FAZIA campaign}
	\begin{figure}[htbp]
		\centering
		\includegraphics[width=0.5\textwidth]{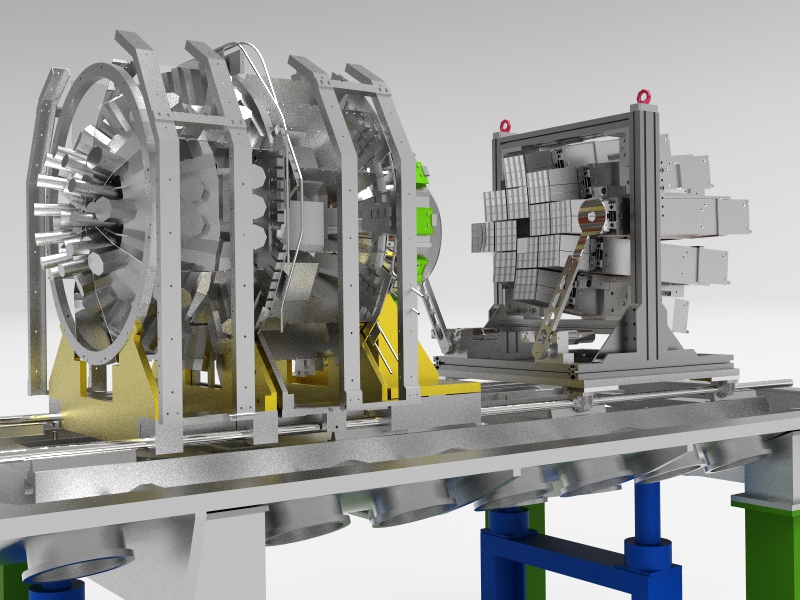}
		\caption{
			INDRA-FAZIA mechanical drawing. It consists of the usual $4\pi$ INDRA multi detector setup, where the first five rings have been removed
			to ensure the backward angular coverage from \SI{14}{\degree} to \SI{176}{\degree} and twelve FAZIA blocks, from \SI{1.5}{\degree} to \SI{14}{\degree},
			divided in four triplets. The whole set-up is hold in the already existing INDRA vacuum chamber at GANIL.
		}
		\label{fig:indrafazia}
	\end{figure}
	The INDRA-FAZIA experimental phase at GANIL is foreseen during the period \numrange{2019}{2023}. It consists of several nuclear physics experiments
	with different beams (e.g. Ca, Ni, Kr, Sn and Xe) at various energies (from 25 to \SI{80}{MeV/u}) on a large variety of targets.
	It will use the coupling between the multi detectors INDRA and FAZIA.
	This set up will be fitted in the already existing INDRA vacuum chamber at GANIL.
	The forward part of the $4\pi$ INDRA multi detector is going to be replaced by the new FAZIA array in order to benefit from its better isotopic,
	energy and angular resolution. The first five INDRA rings (from \SI{1.5}{\degree} to \SI{14}{\degree})
	will be replaced by twelve FAZIA blocks (Fig.~\ref{fig:indrafazia}) as described previously. Those twelve blocks will be divided in four triplets.
	A triplet consists of three FAZIA blocks in a ``L'' configuration. The blocks are mounted inside a metallic structure and fixed by brackets.
	This triplet frame holds many handles, various towing rings and also two removable arms in order to help the mounting
	and the handling inside the vacuum chamber. The main chassis, holding the four triplets, is constituted by a vertical square Elcom based structure,
	where the four angles have been reinforced with thick stainless steel plates. The triplets are then fixed to each four angles of the vertical main square
	frame via screws on the thick plates. An intermediate angled brace between the triplets and each angle of the main square frame fixes
	the geometry of the blocks, i.e. the distance from the target and the polar angles respect to the beam axis where the detectors pointed out.
	For the first experiment at GANIL this intermediate brace has been designed to ensure a one meter distance to the target and a minimum (maximum) polar angles of
	\SI{1.5}{\degree} (\SI{14}{\degree}). The vertical main square frame is eventually screwed to a movable platform which uses the already existing INDRA vacuum chamber rails,
	enabling the right positioning distance to the target. Two remotely controlled thick disk shields, with different diameters
	(a small one protecting the angles from \SI{0.85}{\degree} to \SI{3.1}{\degree} and a big one from \SI{0.85}{\degree} to \SI{5}{\degree}),
	have been designed to protect the silicon detectors during beam focalization or data taking to avoid radiation damage \cite{Barlini13} with stopped heavy ions.
	
	The whole final setup with twelve FAZIA blocks weights around \SI{230}{kg} (a complete single FAZIA block weights slightly less than \SI{15}{kg})
	and it is foreseen to be towed by a crane if necessary. The CAD design and building of the whole support frame for the twelve blocks have been designed at
	the Laboratoire de Physique Corpusculaire de Caen, LPC Caen (France).
	
	\subsection{Cooling}\label{ssc:cooling}
	Since a single FAZIA block absorbs almost \SI{300}{W}, a very efficient cooling solution must be adopted to operate under vacuum.
	The final setup consists in a thick copper plate, on which all the cards are screwed on. The conduction is ensured by thermal grease
	between each card and the copper surface.
	The copper plate has been designed in order to efficiently distribute the liquid flow along the entire surface which holds
	the 8 FEE cards. Solving the conflict between the internal pipes for the liquid flow and the many screw holes
	needed to ensure mechanical and thermal coupling of the electronic boards has been a difficult issue.
	In total, we have about 100 screws. The adopted solution has been to start from a copper slab \SI{8}{mm}
	thick having two main internal in-out pipes running longitudinally on
	the two sides of the plate. The liquid distribution is then ensured in the copper volume through eight transversal
	holes, drilled in the plate and joining the input-output main lateral pipes.
	These transversal holes have variable diameters along the plate length in
	order to compensate for the pressure drop at the various distances from the entrance/exit tube fittings.
	
	Outside the scattering chamber a powerful chiller (ACW LP60) is mounted to refrigerate the water (with \SI{30}{\%} alcohol or glycol) flowing through all the blocks.
	The temperature of the cooling liquid is kept at about \SI{10}{\celsius}.
	To ensure an independent cooling for each block a so called ``clarinet'' device has been built to dispatch the fluid to all the cooling circuits.
	It was designed at the \textit{Grand Accélérateur National d’Ions Lourds} (GANIL).
	
	\subsection{Laser angle measurement}\label{ssc:laser}
	To conclude this review on the FAZIA technological solutions, a notable mention goes to a method that the collaboration implemented to
	precisely measure the polar angles $\vartheta$ and $\varphi$ of the detectors with respect to the beam direction.
	First of all, a piece of beam line beyond the scattering chamber is dismounted to allow the mounting of a laser.
	This laser, aligned with the centre of beam line and the target, is fired in the opposite direction with respect to the beam.
	In place of the target a mirror is then mounted in such a way that its centre is exactly on the path of the laser.
	The mirror equipment is mounted in a gimbal configuration allowing the rotation around two orthogonal axes:
	the inclination and declination can be controlled via two precise stepping motors. The fine regulation of the ``zero'' position
	(when the laser bounces back in the same direction from where it impinged on the mirror) is set once via micrometric screws.
	Then, by regulating the two mirror angles (using the stepping motors) up to when the reflected
	laser beam impinges on the centre of a telescope, it is easy to obtain its $\vartheta$ and $\varphi$ angles
	via a reading of the encoded current position of the motors.
	The centre of the telescope is determined by visual inspection with an estimated accuracy of the order of \SI{1}{mm},
	corresponding to $\sim$\SI{3}{'} accuracy on angle measurement.
	The described technique is in fact a simplified version of a method used by the collaboration for
	the FIASCO experiment (see Sec~2.1 of \cite{Bini03}).
	
	\section{Conclusions and future improvements}\label{sec:conclusions}
	In this review paper we examined the most peculiar and innovative features of the FAZIA telescope array from the technological point of view.
	To summarise, some of the most important characteristics of FAZIA are the compactness and modularity: since FAZIA is structured
	in independent blocks, various geometry configurations are possible; moreover, all the analogue chains operate under vacuum, very close
	to the detectors, so that electronic noise pick-up and signal distortion along the transmission lines are greatly reduced. Another factor that should not be neglected is the optimisation
	of the analogue stages done in the last years: in fact, the pre-amplifiers were designed after many tests performed in the past with
	prototype telescopes \cite{Hamrita04}; a similar consideration is valid also for the choice of the ADCs. The combination of pre-amplifiers which have
	a very high dynamic range and low noise with converters that have a well balanced design in term of ENOB and sampling rate permits
	to obtain a good energy resolution in a very wide spectrum of particle energies and charges (from Am $\alpha$ sources to heavy
	nuclei with energies of the order of \SI{4}{GeV}).
	The clock distribution is another fundamental part of the FAZIA apparatus. Thanks to a very complex design that uses fixed latency
	optical connections and many PLL devices on each electronic card, it is possible to achieve a synchronisation among all the
	acquired channels within \SI{100}{ps}.
	
	Finally, other important aspects of FAZIA are its flexibility and upgradability. In fact, all the electronic boards
	contain programmable devices which are steadily maintained and updated to include new features.
	Indeed, one can implement new firmware solutions on the on-board FPGA after that a given algorithm has been tested
	off-line using the entire sampled waveforms previously acquired.
	For example, in this respect we are going to implement the pulse-shape discrimination of silicon signals directly
	on-board, by adding the search of the current signal maximum on the front-end card FPGA code.
	Another feature which may be added to the same code is an energy shaper on the QL1 signal, to have
	the high gain energy measurement without the need to send the whole signal to the acquisition.
	To improve the data bandwidth, especially in view of experiments that will use many FAZIA blocks, we are also going to implement
	two Ethernet connections (instead of one as we have now) between the regional board and the acquisition system.
	
	This work was partly supported by the Polish Ministry of Science and Higher Education under Contract No. 778N - FAZIA/2010/0,
	the Polish National Science Centre under Contracts No. 2013/08/M/ST2/00257 (COPIGAL) and No. 2014/14/M/ST2/00738 (COPIN-INFN Collaboration).
	
	\bibliography{biblio}
	
\end{document}